\documentclass[twocolumn,showpacs,preprintnumbers,amsmath,amssymb,superscriptaddress]{revtex4}

\usepackage{graphicx}
\usepackage{dcolumn}
\usepackage{bm}

\begin{document}

\preprint{APS/123-QED}

\title{Spatial structure of an individual Mn acceptor in GaAs}

\author{A.M. Yakunin}
 \affiliation{COBRA Inter-University, Eindhoven University of Technology, P.O.Box 513, NL-
5600MB Eindhoven, The Netherlands}
 \author{A.Yu. Silov}
 \affiliation{COBRA Inter-University, Eindhoven University of Technology, P.O.Box 513, NL-
5600MB Eindhoven, The Netherlands}
 \author{P.M. Koenraad}
 \affiliation{COBRA Inter-University, Eindhoven University of Technology, P.O.Box 513, NL-
5600MB Eindhoven, The Netherlands}
\author{J.H. Wolter}
\affiliation{COBRA Inter-University, Eindhoven University of Technology, P.O.Box 513, NL-
5600MB Eindhoven, The Netherlands}
\author{W. Van Roy}
\affiliation{IMEC, Kapeldreef 75, B-3001 Leuven, Belgium}
\author{J. De Boeck}
\affiliation{IMEC, Kapeldreef 75, B-3001 Leuven, Belgium}

\author{J.-M. Tang}
\affiliation{Optical Science and Technology Center and Department of Physics and Astronomy,
University of Iowa, Iowa City, IA 52242, USA}
\author{M.E. Flatt\'{e}}
\affiliation{Optical Science and Technology Center and Department of Physics and Astronomy,
University of Iowa, Iowa City, IA 52242, USA}

\date{\today}

\begin{abstract}
The wave function of a hole bound to an individual Mn acceptor in GaAs is
spatially mapped by scanning tunneling microscopy at room temperature and an
anisotropic, cross-like shape is observed. The spatial structure is compared with that from
an envelope-function, effective mass model, and from a tight-binding model. This
demonstrates that anisotropy arising from the cubic symmetry of the GaAs crystal
produces the cross-like shape for the hole wave-function. Thus the coupling between Mn
dopants in GaMnAs mediated by such holes will be highly anisotropic.
\end{abstract}

\maketitle
Despite intense study of deep acceptors in III-V semiconductors such as Mn$_{\text{Ga}}$,
little information has been obtained on their electronic properties at the atomic scale. Yet
the spatial shape of the Mn acceptor state will influence hole-mediated Mn-Mn coupling
and thus all of the magnetic properties of hole-mediated ferromagnetic
semiconductors \cite{a1} such as Ga$_{1-x}$Mn$_{x}$As. Evidence for anisotropic spatial structure even
in shallow acceptors such as Zn, Cd \cite{Zn,ZnCd} and C \cite{Carbon} in various III-V semiconductors
suggest that anisotropic hole states may be common. In addition to controlling inter-dopant properties such
as the Mn-Mn coupling \cite{a3,Tang}, the wave-function shape would affect single-dopant properties such as the g-factor and optical transition oscillator strengths.

This study presents an experimental and theoretical description of the spatial
symmetry of the Mn acceptor wave-function in GaAs, and we suggest our results imply
similar behavior for other acceptors and other hosts. We first present our measurements
of the spatial mapping of the anisotropic wave function of a hole localized at a Mn
acceptor. To achieve this, we have used the STM tip not only to image the Mn acceptor
but also to manipulate its charge state $A^{0}/A^{-}$ at room temperature as described in \cite{Manipulation}. Within an envelope
function effective mass model (EFM) the anisotropy in the acceptor wave-function can be
traced to differing amplitudes of envelope functions with the same total angular
momentum ($L>0$) but different angular momentum projections along a fixed axis. We introduce into the EFM a single parameter $\eta$ that describes the breaking, by the cubic crystal, of spherical symmetry for the acceptor level envelope functions. As $\eta$ has a negligible effect on the binding energy
compared to the central cell correction, common variational approaches cannot be used to
evaluate $\eta$. However, comparison with calculations based on a tight-binding model
(TBM) for the Mn acceptor structure \cite{Tang} permits us to clearly identify the physical origin
of the anisotropic shape in these models, and to justify the value of $\eta$ used in our EFM
calculations to describe the experimental shape. The TBM calculations also demonstrate
that although the spin-orbit interaction does influence the acceptor wave-function, the
qualitative anisotropic shape of the acceptor state occurs in crystals without spin-orbit
interaction. Thus acceptor levels in crystals such as GaN should have a similar shape.

The measurements were performed on several samples using chemically etched
tungsten tips. The samples consisted of a 1200~nm thick layer of GaAs doped with Mn at
$3\times10^{18}$~cm$^{-3}$ grown by MBE on an intrinsic (001) GaAs substrate. A growth temperature
of 580~$^{\text{o}}$C was chosen to prevent the appearance of structural defects such as As antisites,
which would complicate the spatial mapping by shifting the position of the Fermi level
of the sample. The concentration of the Mn dopants was low enough to neglect Mn-Mn interactions and the formation of an impurity band. The samples we used were insulating below 77~K. The experiments were performed in a room temperature UHV-STM
($P<2\times10^{-11}$~torr) on an \emph{in situ} cleavage induced (110) surface.
\begin{figure}[t]
  \includegraphics{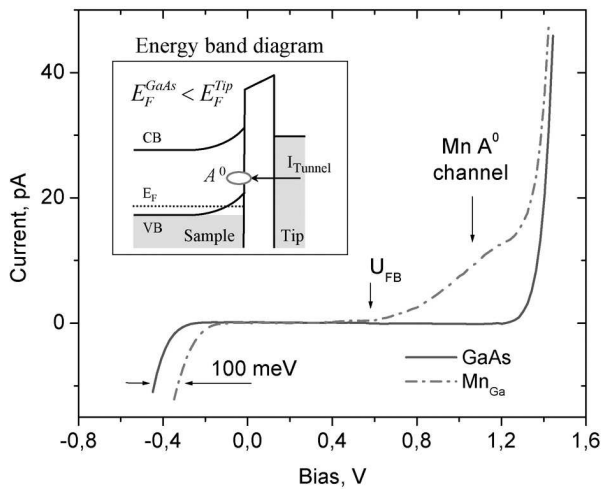}
  \caption{I(V) curves acquired on the clean GaAs surface (solid line) and in
the middle of the cross-like feature(dashed line). The simulated position of the flat-band potential $U_{FB}$ is
indicated by the arrow. Insert displays energy band diagram for the positive sample-bias.}\label{IV}
\end{figure}

\begin{figure}
  \includegraphics{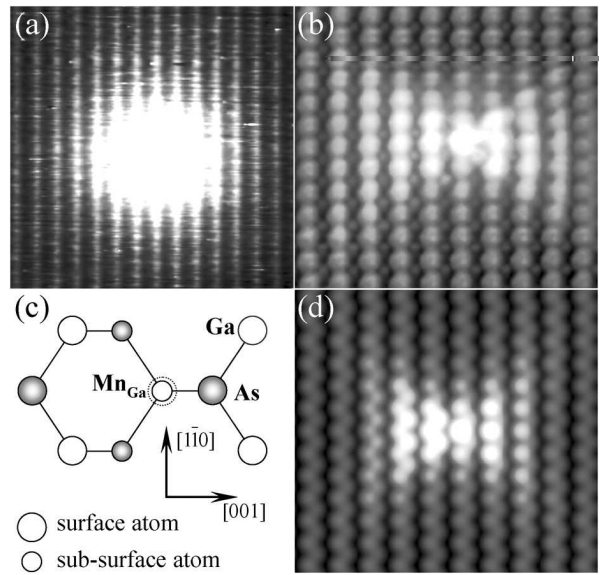}
  \caption{(a)~8x8~nm$^{2}$ STM image of an ionized Mn acquired at -0.7~V; (b)~5.6x5~nm$^2$ STM
image of neutral Mn acquired at +0.6~V. Big and small round features correspond to As
and Ga related surface states, respectively. Presumably, Mn is located in 3$^{rd}$ sub-surface
atomic layer; (c)~A model of the (110)-surface (top view) representing the Mn on Ga site
located in an odd sub-surface atomic layer (counting surface layer as zero); (d)~5.6x5~nm$^2$
simulated image (logarithm of local density of states) of the Mn located in the 5$^{th}$ sub-surface atomic layer (TBM).}\label{Atomic}
\end{figure}

\begin{figure}
  \includegraphics{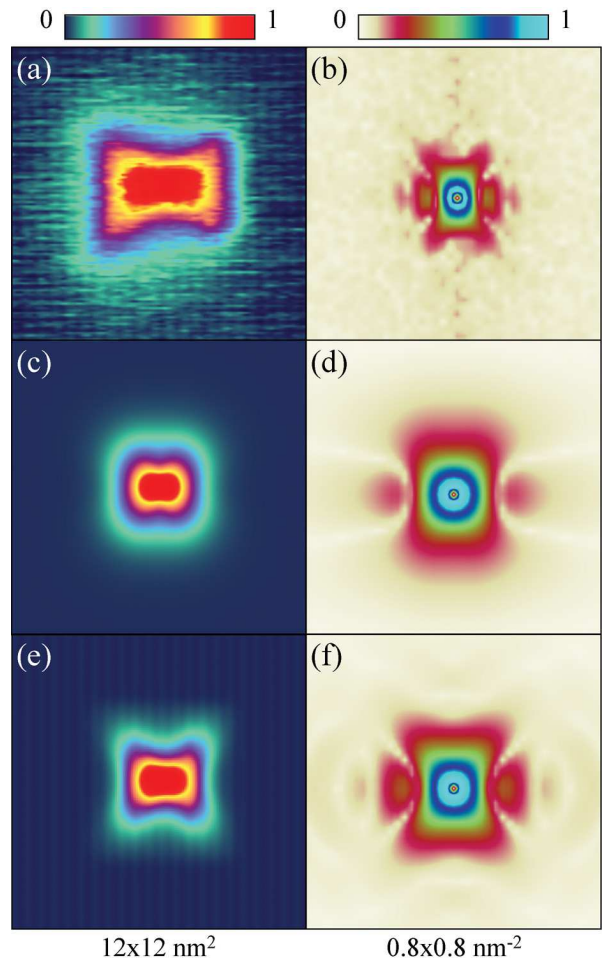}
  \caption{(a)~Constant-current image of the neutral Mn acquired at +0.9~V. Mn is presumably
located in the $2^{nd}$ sub-surface atomic layer; (b)~Fourier spectrum of the image (a);
(c)~simulated image of the Mn acceptor ground state $1S_{3/2}$ (EFM); (d)~Fourier spectrum of
the image (c); (e)~simulated image of the Mn acceptor DOS (TBM); (f)~Fourier spectrum of
the image (e). Images (c) and (e) were simulated assuming Mn to be located in the $4^{th}$
sub-surface atomic layer.}\label{Fourier}
\end{figure}

A major advantage to our approach is that the occupation of the acceptor state can
be influenced by band-bending from the voltage applied between the STM tip and the
sample (see the Fig.~\ref{IV} inset). We studied the voltage dependent appearance of the Mn
acceptor in the STM constant-current image. In the ionized configuration at high negative
bias Mn appeared as an isotropic round elevation which is a consequence of the influence
of the $A^{-}$ ion Coulomb field on the valence band states [Fig.~\ref{Atomic}(a)]. This agrees with a
recent study of the individual Mn in GaAs in the ionized configuration \cite{MnIonized}. We found that
at a positive bias Mn is electrically neutral, as can be seen from the absence of the
electronic contrast at high positive voltage ($U>1.5$~V) when the conduction band empty
states dominate tunneling (see the positive branch of the $I(V)$ curve in Fig.~\ref{IV}). At low
positive voltage where the tip Fermi level is below the conduction band edge Mn
appeared as a highly anisotropic cross like feature [Figs.~\ref{Atomic}(b) and \ref{Fourier}(a)]. The anisotropy is
even more evident in a reciprocal-space image [Fig.~\ref{Fourier}(b)]. The methods of calculating the
theoretical images [Figs.~\ref{Atomic}(d),~\ref{Fourier}(c-f)] will be described below.

The cross like feature manifested itself in the local tunneling $I(V)$ spectroscopy at
low voltages when the GaAs bands do not contribute to the tunneling. It appeared as an
empty states or filled states current channel in the band gap of GaAs depending on
applied positive or negative bias, respectively. Thus the mapping of the Mn acceptor state
in the filled (empty) states mode was realized by electron (hole) injection into the $A^{0}/A^{-}$~
state. In the tunneling $I(V)$ spectroscopy the manganese $A^{0}$ channel appeared presumably
above the flat band potential $U_{FB}$ and was available for tunneling in the wide range of
voltages above $U_{FB}$. Our estimated value of $U_{FB}$ is about +0.6~V. The observed ionization
energy, which was determined from the shift of the $I(V)$ spectrum at negative bias,
corresponds to the Mn acceptor binding energy $E_{a}=0.1$~eV.

The concentration of the dopants we observed with STM corresponds to the
intentional $3\times10^{18}$~cm$^{-3}$ doping level. All of the dopants could be found either in the
ionized $A^{-}$ or the neutral $A^{0}$ charge state depending on either negative or positive sample
bias respectively. In the experiment we identified Mn located in at least 6 different layers
under the surface. In order to determine the actual position of the Mn dopants we
analyzed the intensity of the electronic contrast of the Mn related features. Based on the
symmetry of the cross like feature superimposed on the surface lattice we distinguished
whether the dopant is located in an even or odd sub-surface layer. We found that at any
depth the shape of the hole on \{110\} surface had nearly $C_{2v}$ symmetry around the surface normal, and was elongated along the [001] direction relative to the [1$\bar{1}$0] direction. The cross-like features induced
by deeper dopants were more elongated in the [001] direction. The feature was weakly
asymmetric with respect to [1$\bar{1}$0] direction (lowering the symmetry to the single (1$\bar{1}$0)
mirror plane) which may come from the symmetry properties of the (110) surface.

A four-band envelope-function effective Luttinger-Kohn Hamiltonian provided
one framework (EFM) in which to analyze the spatial structure of the bulk-like neutral
acceptor complex formed by a valence hole loosely bound to a negatively charged
Mn$^{2+}3d^{5}$ core (Mn$^{2+}3d^{5}+\text{hole}$~complex). The ground state of this acceptor in a zincblende
semiconductor can be approximated as four-fold degenerate with a total
momentum of the valence hole $F=3/2$ and has the symmetry of the top of the valence
band $\Gamma_{8}$ \cite{Pikus}. We neglected possible effects caused by the presence of the (110) surface
and quantum spin effects from the exchange interaction between the Mn$3d^{5}$ core and the
hole. We also ignored the excited states, as the energy separation between the ground
state $E_{a}(1S_{3/2})=113$~meV and the first excited state $E_{a}(2S_{3/2})=25$~meV exceeds room
temperature \cite{Ea}.

According to Ref.~\onlinecite{Pikus} the acceptor wave-function in zincblende
semiconductor is represented as a four-component column written in the basis of Bloch
functions of the valence band top $\Gamma_{8}$. The form of the wave-function component $\psi^{3/2}_{3/2}(\theta,\varphi,r)$ is proportional to:
\begin{equation}\label{Krasota}
R_{_{0}}\!
\left|\!\left|\!\!
\begin{array}{c}
  Y_{_{0,0}} \\
  0 \\
  0 \\
  0 \\
\end{array}
\!\!\right|\!\right|
+c^{F_{1}}R_{_{2}}\!
\left|\!\left|\!\!
\begin{array}{c}
  0 \\
  2Y_{_{2,1}} \\
  Y_{_{2,2}}\!\!-\!Y_{_{2,-2}} \\
  0 \\
\end{array}
\!\!\right|\!\right|
+c^{E}R_{_{2}}\!
\left|\!\left|\!\!
\begin{array}{c}
  \sqrt{2}Y_{_{2,0}} \\
  0 \\
  Y_{_{2,2}}\!\!+\!
  Y_{_{2,-2}} \\
  0 \\
\end{array}
\!\!\right|\!\right|,
\end{equation} 
\begin{equation}\label{EquEta}
\text{where }
c^{E}/c^{F_{1}}=\eta, \text{ }0 \leq\eta\leq1,
\end{equation}
$Y_{L,m}(\theta,\varphi)$ are spherical harmonics, $R_{L}(r)$ are radial parts of the envelope
functions, $c^{E}$ and $c^{F_{1}}$ are constants described below.

As pointed out by Kohn and
Schechter \cite{Schechter,Kohn} in this model $R_{L}(r)Y_{L,m}(\theta,\varphi)$ are eigenfunctions of parity. Thus the
wave function [Eq.~(\ref{Krasota})] contains only even s-like ($L=0$) and d-like ($L=2$) components. In
Eq.~\ref{Krasota} the angular part of the d-component includes components which transform
according to the $\Gamma_{12}$ and $\Gamma_{25}$ representations of the tetrahedral point group. Their
corresponding coefficients are the constants $c^{E}$ and $c^{F_{1}}$, respectively, whose ratio is
denoted here as $\eta$ [Eq.~\ref{EquEta}].
The value of $\eta$ depends on the valence band parameters of the particular material
and could be evaluated variationally from the binding energy \cite{Pikus,Schechter,Kohn}. However the
introduction of $\eta$ only weakly affects the binding energy and is negligible compared to a
central cell correction. To clarify the nature of the Mn acceptor ground state here we used
$\eta$ as a free parameter, and thus varied the spherical anisotropy from $\eta=1$ , corresponding
to a completely spherically symmetric solution, till $\eta\rightarrow0$ , where the E component
vanished. When $\eta=1$, the function [Eq.~\ref{Krasota}] transforms to that one obtained by Baldereschi and Lipari in the spherical approximation~\cite{Lipari,Baldereschi}.

Although the EFM describes the symmetry of
the Mn acceptor, it fails to predict the ground-state binding energy $E_{0}$ without a central
cell correction. Deep acceptors such as Mn in GaAs are usually described by the zero-range
potential model \cite{Perel,ZeroRange}. In this model the spatial extension of the wave function
depends on a parameter $\alpha$, determined by $E_{0}$:
\begin{equation}\label{AlphaBeta}
    \alpha=\frac{\sqrt{2m_{h}E_{0}}}{\hbar}\left(\frac{\beta\sqrt{\beta}+1}{\beta+1}\right),\text{ }
    \beta=\frac{m_l}{m_h},
\end{equation}
where the parameter $\alpha$ characterizes the spatial localization of the hole trapped at the
acceptor and $\beta$ is the ratio of light to heavy hole masses. For GaAs we used $\alpha=1.197$~nm$^{-1}$ and $\beta=0.132$. This model allows one to obtain analytical expressions for $R_{0}(r)$ and
$R_{2}(r)$ \cite{ZeroRange}. The admixture of the d-like component to the ground state calculated by
Baldereschi and Lipari \cite{Baldereschi} for GaAs is 30\% and corresponds to what
we have found using the zero-range potential model. As can be seen from Ref.~\onlinecite{ZeroRange} the d-like
component dominates at distances $R>1/\alpha$. The admixture of the d-like
component and thus the spatial structure of the confined hole are strongly dependent
on $\beta$. The admixture of higher orbital momentum will vanish in the limit $\beta\rightarrow1$. In the
case $\beta=0$ the envelope functions $R_{0}$ and $R_{2}$ can be derived in an implicit form \cite{Linnik}. In
Ref.~\onlinecite{Linnik} authors emphasize the different behavior of $R_{0}$ and $R_{2}$ at large distances. The $R_{0}$ drops exponentially whereas $R_{2}$ has power law decay at infinity.

In Fig.~\ref{Fourier}(c) we present the cross-section of the bulk-like charge density of the
total wave function in the limit case $\eta=0$ cut through an imaginary (110) plane at the
distance of 0.8~nm from the Mn position. The plot is presented with a logarithmic scale
because of the inverse exponential dependence of the tunnel current on the tip-sample
distance during the actual STM experiment. Figure~\ref{Fourier}(d) presents the calculated Fourier
transform. Note the presence of the satellite harmonics that arise from the steep fall off of
the wave function in the [001] direction. The model reproduces the key symmetry
elements in the Fourier spectra: the elongation of the spectra along [1$\bar{1}$0] direction, the
satellites in [001] direction and the cross like shape. However the cross-like shape and the
intensity of the satellites are more pronounced in the experimental Fourier spectrum.
Based on the analyses we performed we conclude that only the harmonic $Y_{2,1}$ gives rise to
the intensity of the satellites in the Fourier spectrum. The cross feature in the Fourier
spectrum is also influenced by the envelope function $R_{2}$ and is more pronounced when $R_{2}$
has inverse power dependence on distance rather then exponential decay.

The tight-binding model (TBM) we use \cite{Tang} is based on the deep level model of
Vogl and Baranowski \cite{Vogl}. The dangling $sp^{3}$-bonds from the nearest-neighbor As
hybridize with the Mn d-states of $\Gamma_{15}$ character. The antibonding combination of these
becomes the Mn acceptor state. Coupling to the d-states of $\Gamma_{12}$ character is weak, and
hence neglected. The hybridization strength is fully determined by the acceptor level
energy. This model, if further approximated within the EFM, predicts $\eta=0$, similar to
what we found by fitting the EFM to the experimental measurements.

The calculations of the local density of states (LDOS) based on the TBM are shown in logarithmic scale in Fig.~\ref{Atomic}(d) and Figs.~\ref{Fourier}(e,f). The results show symmetry under reflection in the (1$\bar{1}$0) plane, and asymmetry
under reflection in the (001) plane. For Mn dopants several layers down from the surface,
as in Fig.~\ref{Atomic}(b), the shape of the acceptor state does not depend that sensitively on the spin
orientation of the Mn-core $3d$-spin. For these dopants the shape does not depend on the spin-orbit
interaction; we confirmed this by obtaining a similar cross-like acceptor structure using a tight-binding Hamiltonian without spin-orbit interaction, and with empirical parameters designed for optimal 
agreement with the bulk band structure of GaAs~\cite{NonSpinOrbit}. The situation differs greatly
for Mn near the surface, where the axis of extension of the acceptor state rotates with the
spin orientation of the Mn core spin. At these temperatures we can expect the Mn core
spin to point in a random direction, and for the STM measurements to average over the
possible spin orientations. Thus for Figs.~\ref{Atomic}(d) and \ref{Fourier}(e,f) the LDOS is averaged over the
Mn spin orientation.

Although the symmetry is well retained for all simulated positions of Mn under
the surface, in both models the best fits were achieved when the apparent depth of the Mn
is assumed to be 2 atomic layers larger. The reason for this could be the vacuum barrier,
which will tend to shift the wave function of the Mn acceptor state deeper into the crystal
than one would predict for the sliced bulk crystal calculation. The lateral size of the wave
function measured is also somewhat larger than calculated for Fig.~\ref{Fourier}; this might come
from a reduction of the acceptor binding energy very near the surface.

In conclusion, we have experimentally demonstrated that the Mn acceptor ground
state has highly anisotropic spatial structure. This spatial anisotropy is due to a significant
presence of d-wave envelope functions in the acceptor ground state. We have demonstrated that the
observed symmetry can be explained within simple tight-binding model whose only free
parameter is the acceptor level energy. We also found that this spatial structure can be
described well by a simple four-band envelope-function model of cubic symmetry whose
key parameter, $\eta$, can be fit to the observed spatial structure, and is similar to that
expected from a tight-binding model. These results have broad implications for all
acceptor-acceptor interactions in zincblende semiconductors, and especially for hole-mediated
ferromagnetic semiconductors.

This work was supported by the Dutch Foundation for Fundamental Research on Matter
(FOM), the Belgian Fund for Scientific Research Flanders (FWO), the EC GROWTH
project FENIKS (GR5D-CT-2001-00535), and the ARO MURI DAAD-19-01-1-0541.


\begin{thebibliography}{}
\bibitem[Ohno (2002)]{a1} H. Ohno, \emph{in Semiconductor Spintronics and Quantum Computation,}
edited by D.D. Awschalom, N. Samarth, and D. Loss (Springer, Berlin, 2002).
\bibitem[Zheng (1994)]{Zn} J.F. Zheng, M. Salmeron, and E.R. Weber, Appl. Phys. Lett. \textbf{64}, 1836 (1994); \textbf{65}, 790(E) (1994).
\bibitem[Kort (2001)]{ZnCd} R. de Kort, M. C. M. M. van der Wielen, A. J. A. van Roij, W. Kets, and H. van
Kempen, Phys. Rev. B \textbf{63}, 125336 (2001). 
\bibitem[Reusch (2003)]{Carbon} T.C.G. Reusch, M. Wenderoth, L. Winking, R.G. Ulbrich, G. Döhler, S. Malzer
in \emph{Proceedings of the STM'03 conference} (Eindhoven, 2003).
\bibitem[Zarand (2002)]{a3} G. Zar\'{a}nd and B. Jank\'{o}, Phys. Rev. Lett. \textbf{89}, 047201 (2002).
\bibitem[Tang (2004)]{Tang} J.-M. Tang and M. E. Flatt\'{e}, Phys. Rev. Lett. \textbf{92}, 047201 (2004).
\bibitem[Yakunin (2004)]{Manipulation} A.M. Yakunin, A.Yu. Silov, P.M. Koenraad, W. Van Roy, J. De Boeck, J.H. Wolter, Physica E in press.
\bibitem[Tsuruoka (2002)]{MnIonized} T. Tsuruoka, R. Tanimoto, N. Tachikawa, S. Ushioda, F. Matsukura and H. Ohno,
Solid State Com. \textbf{121}, 79 (2002).
\bibitem[Pikus (1974)]{Pikus} G.L. Bir and G.E. Pikus in \emph{Symmetry and strain-induced effects in semiconductors},
(Halsted, Jerusalem, 1974).
\bibitem[Linnarsson (1997)]{Ea} M. Linnarsson, E. Janzén, B. Monemar, M. Kleverman, and A. Thilderkvist, Phys.
Rev. B, \textbf{55}, 6938 (1997).
\bibitem[Schechter (1962)]{Schechter} D. Schechter, J. Phys. Chem. Sol. \textbf{23}, 237 (1962).
\bibitem[Kohn (1955)]{Kohn} Kohn W. and D. Schechter, Phys. Rev. \textbf{99}, 1903 (1955).
\bibitem[Lipari (1970)]{Lipari} N.O. Lipari and A. Baldereschi, Phys. Rev. Lett. \textbf{25}, 1660 (1970).
\bibitem[Baldereschi (1973)]{Baldereschi} A. Baldereschi and N.O. Lipari, Phys. Rev. B \textbf{8}, 2697 (1973).
\bibitem[Perel (1982)]{Perel} V.I. Perel' and I.N. Yassievich, Zh. Eksp. Teor. Fiz. \textbf{82}, 237 (1982) [Sov. Phys. JETP \textbf{55}, 143 (1982)].
\bibitem[Averkiev (1982)]{ZeroRange} N.S. Averkiev and S. Yu. Il'inskii, Phys. Solid. State \textbf{36}(2), 278 (1994) [Fiz. Tverd. Tela, \textbf{36}, 503 (1994)].
\bibitem[Sheka (1999)]{Linnik} T.L. Linnik and V.I. Sheka, Phys. Solid State \textbf{41}, 1425 (1999).
\bibitem[Vogl (1985)]{Vogl} P. Vogl and J. M. Baranowski, Acta. Phys. Polo. \textbf{A67}, 133 (1985).
\bibitem[Chadi (1975)]{NonSpinOrbit} D.J. Chadi and M. L. Cohen, Phys. Status Solidi B \textbf{68}, 405 (1975).

\end{thebibliography}

\end{document}